\documentclass[showpacs,preprintnumbers, twocolumn, amsmath,amssymb]{revtex4}
\usepackage{graphicx}
\usepackage{dcolumn}
\usepackage{bm}
\usepackage{subfigure}

\begin{document}

\title{The Euclid space-time diagram of the theory of relativity}
\author{W. LiMing}\email{wliming@scnu.edu.cn}
\affiliation{Dept. of Physics, and Laboratory of Quantum Information Technology, School
of Physics and Telecommunication Engineering, South China Normal University, Guangzhou
510006, China}
\begin{abstract}A conventional space-time diagram is $r-ct$ one, which satisfies the Minkowski geometry. This geometry conflict the intuition from the Euclid geometry. In this work an Euclid space-time diagram is proposed to describe relativistic world lines with an exact Euclid geometry. The relativistic effects such as the dilation of moving clocks, the contraction of moving length, and the twin paradox can be geometrically expressed in the Euclid space-time diagram. It is applied to the case of a satellite clock to correct the gravitational effect. It is found that this Euclid space-time diagram is much more intuitive than the conventional space-time diagram.
\end{abstract}
\date{\today}
\maketitle

\section{Introduction}

A conventional space-time diagram (STD) is a ${\bf r}-ct$ diagram\cite{hans}. In general a STD takes a two dimensional space and a one dimensional time to create a three dimensional diagram.  A world line in a STD records the positions of a particle at different time, i.e. the track of a particle. An example of a world line is shown in Fig.1, where a particle starts from position ${\bf r_0}$.  A point on a world line corresponds to an event of a process.


A world line in a STD satisfies the Minkowski geometry, which is very different from the Euclid geometry, thus to be easily misleading new learners.    
 
The time recorded by a clock moving together with a particle is called the proper time. It is an invariant quantity under transformations of inertial references. The proper time of an infinitesimal process is defined by $d\tau$ in the following equation
\begin{align}c^2d\tau^2= c^2dt^2 - |d{\bf r}|^2 \label{dt}\end{align}
where $|d{\bf r}|^2 = dx^2 + dy^2 + dz^2$. Quantities $d\tau, dt, |d{\bf r}|$ are shown by a triangle in Fig.1. These quantities, however, do not satisfy the Euclid geometry since they satisfy \eqref{dt}. Alternatively this geometry is called the {\it Minkowski geometry}. Thus the right triangle in Fig.1 is very misleading to new learners, i.e. $cd\tau$ is not the real length of the world line.

\begin{figure}
\subfigure{\includegraphics[width=4cm,height=5cm]{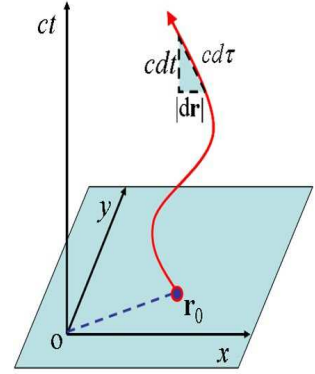}}
\caption{ A world line in the conventional space-time diagram. A particle starts from ${\bf r}_0$ at $t=0$. 
}
\end{figure}

Due to the disadvantage of the conventional STD it is worthy to find a new diagram which is fully geometric to shown the world lines of particles. 

\section{Euclid space-time diagram}

Since in a flat space-time one has
\begin{align} \label{tri} c^2dt^2 = c^2d\tau^2 + |d{\bf r}|^2 \end{align}
the proper time $cd\tau $ and the displacement $|d{\bf r}|$ sustain the two right angle edges of a right triangle, and the coordinate time $cdt$ spans the triangle as the hypotenuse. This is just the case in a ${\bf r}-c\tau$ diagram as shown in Fig.2. It is fully geometric, hence is called an {\it Euclid space-time diagram}(eSTD).
The angle $\theta$ of the right triangle  gives the speed of the particle
\begin{align}
 \cos\theta = |d{\bf r}|/cdt = v/c \end{align}
{\it The geometric length of a world line is exactly equal to} $c\Delta t$. The time interval $\Delta t$ between two events $P$ and $Q$ is given by
\begin{align}
\Delta t &= \int_P^Q dt = \int_P^Q d\tau \sqrt{1 + \cot^2\theta}
 \end{align}

Therefore, the present eSTD is much more intuitive than the conventional STD.

\begin{figure}\label{spacetime}
\includegraphics[width=4cm,height=5cm]{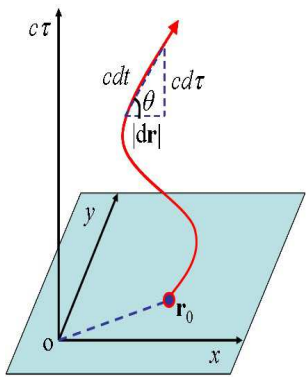}
\caption{A world curve in a geometric space-time diagram.}
\end{figure}

Since a light ray has a zero proper time, $d\tau = 0$, it lies on the $xy$ plain, as shown by the dashed line in Fig.2. Then the light cone in the conventional STD becomes a light plain (i.e. the $xy$ plain) in the present eSTD.

For convenience, the time zero $t=0$ of any world line can be defined on the $xy$ plain. Of course this definition is reference dependent. 

\section{A moving clock and a moving length}

Consider a moving clock in a spacecraft starting from the origin with velocity $u$. The world lines of the moving clock (blue) and a rest clock at the origin (red) in the eSTD are shown in Fig.3(a). At time $t$ the rest clock and the moving clock arrive at a circle of radius $ct$. The time elapsed on the reference of the spacecraft, i.e. the proper time $\tau_2$, is given by the geometry of the blue world line:
\begin{align}
\tau_2 &= t \sin\theta= t \sqrt{1-u^2/c^2} <t
 \end{align}
Therefore, the time elapses slower in a moving reference by a factor of $\sqrt{1-u^2/c^2}$. The is just the time dilation of a moving reference. Again we see that  the eSTD becomes much more straight forward.

The twin paradox is a famous argument in the history of the theory of relativity. Two twin brothers A and B were born at Sanya village some years ago, but brother B was taken away by a high-speed spacecraft of a group of aliens just after he was born. Today the spacecraft comes back. The two twin brothers meet at the place where there were born. According to the special theory of relativity brother B should be younger than A on view of the people of the village. On view of the aliens, however, brother A should be younger than B since they feel that the earth relatively moves away and comes back to them. Who is younger in fact? This is the twin paradox. Of course, physics has given a certain conclusion on this paradox that the view of the village people is correct since the reference of the aliens is not inertial.
\begin{figure}
\subfigure{\includegraphics[width=4cm,height=4cm]{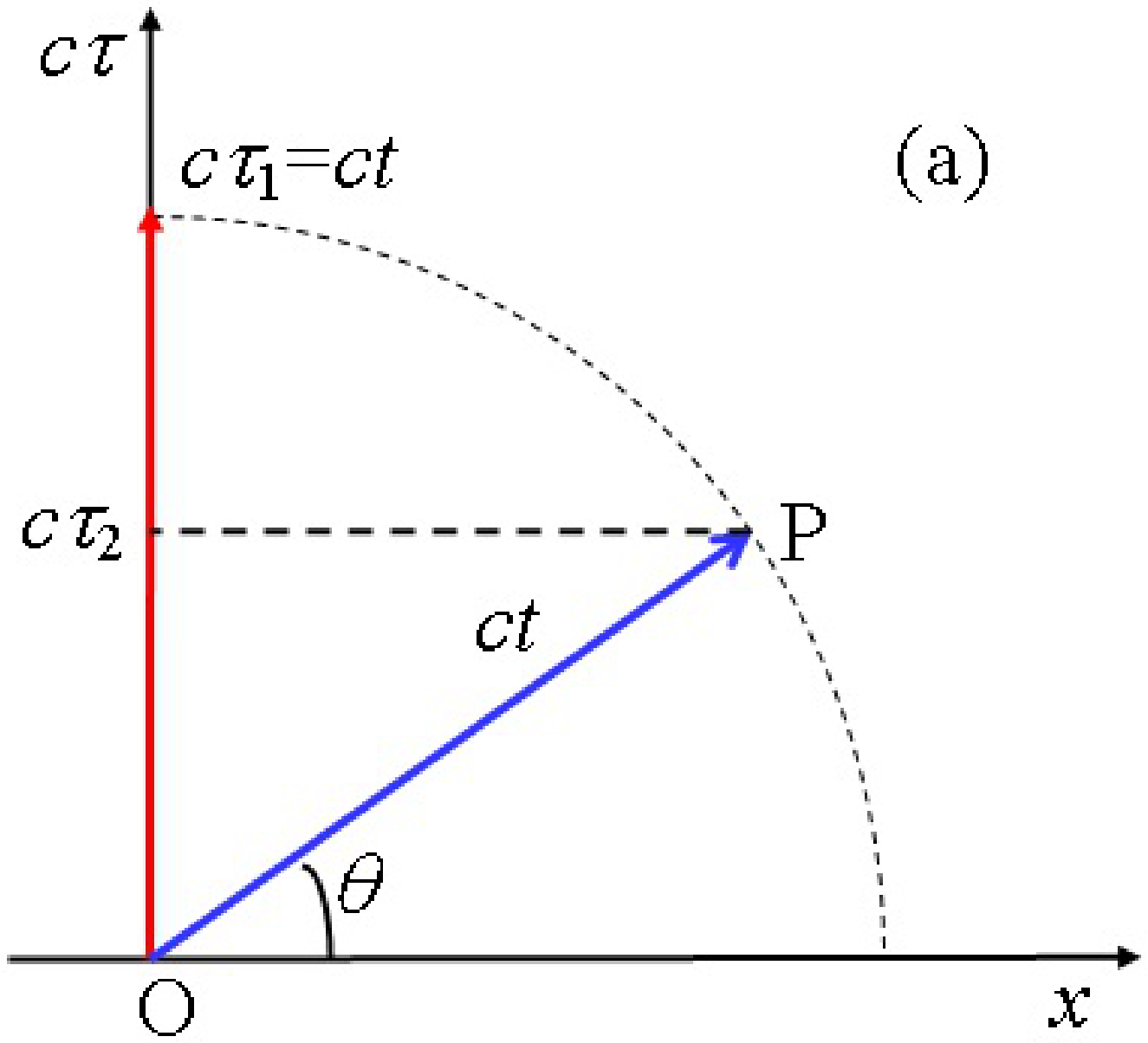}}
\subfigure{\includegraphics[width=4cm,height=4cm]{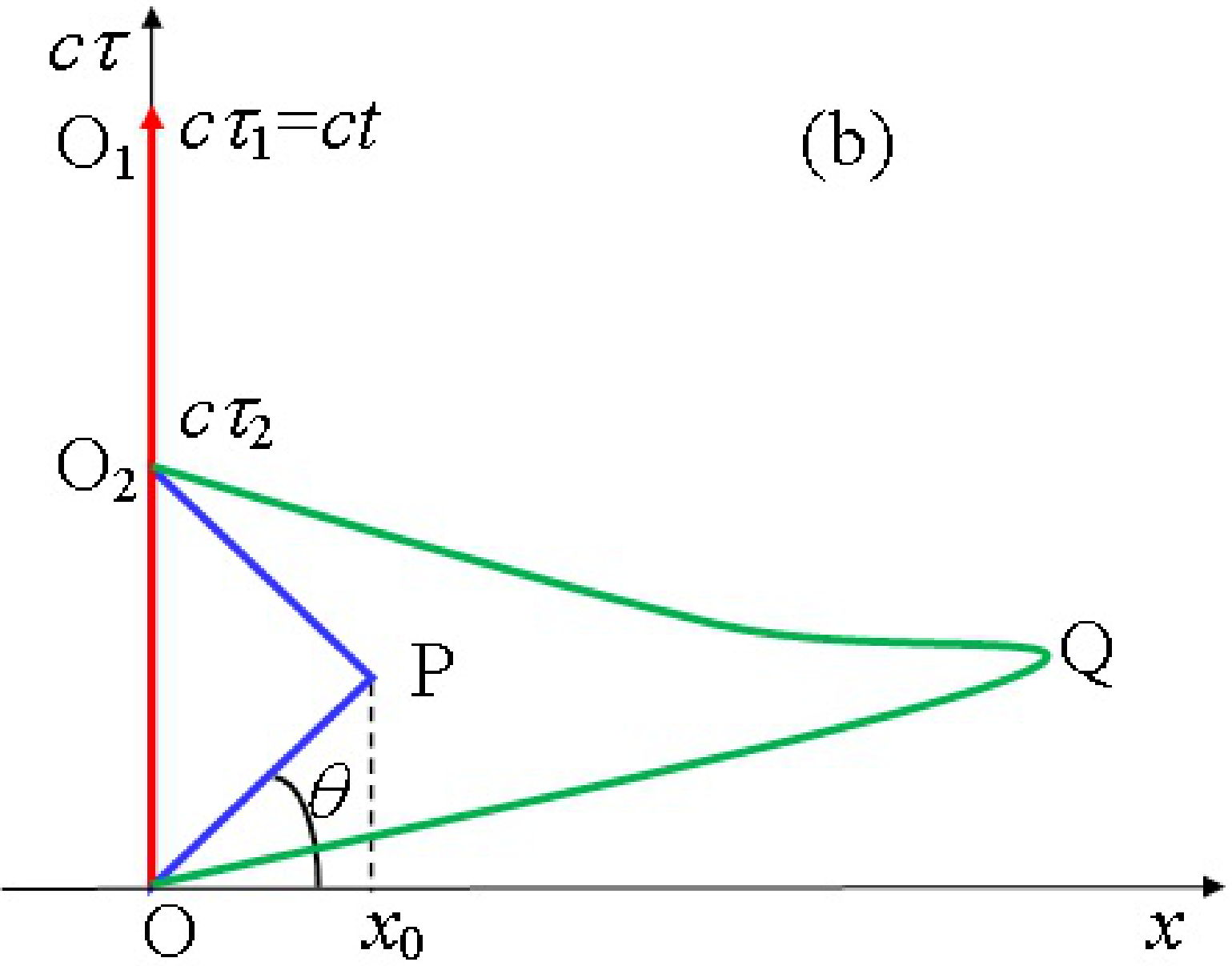}}
\caption{(a) World lines of a moving clock (blue) and a rest clock (red); (b) World lines of two twin brothers(red and blue).
}
\end{figure}
As shown in Fig.3(b) two twin brothers were born at $O$ and $t=0$, and  ${OO_1}$ (red) is the world line of brother A and ${OPO_2}$ (blue) the world line of brother B. B traveled to a distance $x_0$ away from the earth with a constant speed $v$ and comes back to meet A at time $t$. The world lines of the two brothers have the same lengths $ct$ (red and blue). The angle $\theta$ gives the speed of the spacecraft, $\cos\theta = 2x_0/ct = u/c$. The proper time $\tau_2$ is the real time that B experienced. According to the geometry of the eSTD the proper time is given by
\begin{align}
\tau_2 = 2\times {t\over 2} \sin\theta  = t\sqrt{1-{u^2/ c^2}}< t
\end{align}
 Thus brother B is younger than A.   If B traveled with a higher speed, e.g., along $OQO_2$ in Fig.3(b) with a longer path (i.e. a longer time $t'$), but comes back with the same proper time $\tau_2$, A would be an age of $t'$, further older than the former case. Obviously, this analysis is much easier than the conventional STD.


Now we consider a moving length $L_0$ in a reference frame ${\rm S'}$ with velocity $u$ relative to a rest frame ${\rm S}$ as shown in Fig.4. We measure the positions of the two ends of the moving length at time $t=0$ in frame S to given the length $L = x_2-x_1$ (blue bar, $x_1=0$). This happens at different time $t'$ in frame ${\rm S'}$ as shown. Since the length is at rest in ${\rm S'}$ the two ends at any time give $x'_2-x'_1 = L_0$ (green bar).

\begin{figure}
\subfigure{\includegraphics[width=4cm,height=4cm]{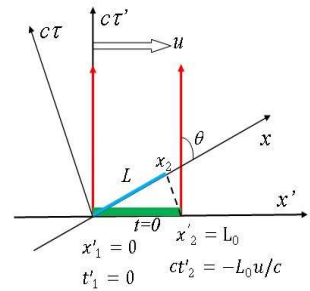}}
\caption{World lines of a moving length $L_0$. 
}
\end{figure}

The geometry of the right triangle in the figure gives
\begin{align}
L&=L_0 \sin\theta \nonumber \\
 &= L_0 \sqrt{1-u^2/c^2}
\end{align}
It is seen that a moving length looks shorter. Again  the eSTD gives intuitively a geometric picture of relativistic effects.

\section{ESTD for GPS system}

This eSTD can be applied to a satellite clock (gravity is not considered at the moment). A satellite is rotating around the earth. The world line of the satellite is shown in Fig.5(a). The cylindrical surface in the diagram contains the world lines of the earth surface. The length of the world line of the satellite is equal to the height of the world line of the earth($ct$), hence the satellite experienced shorter proper time ($c\tau$). According to the right triangle relation (\ref{tri}) of the eSTD one obtains the proper time for each cycle of rotations of the satellite
\begin{align}
c^2\Delta\tau^2 & = c^2T^2 - (2\pi r)^2=(c^2 - v^2)T^2\label{sat}
\end{align}
where $T$ is the period of one cycle of the satellite's rotation. Another relation comes from Newton's second law on the centripetal force ${GM/r^2} = v^2/r$, here $M$ and $r$ are the mass of the earth and the orbital radius of the satellite, respectively. Then one has
\begin{align}
\Delta\tau & =\sqrt{1 - {GM\over rc^2}}T
\end{align}
\begin{figure}
\subfigure{\includegraphics[width=4cm,height=5cm]{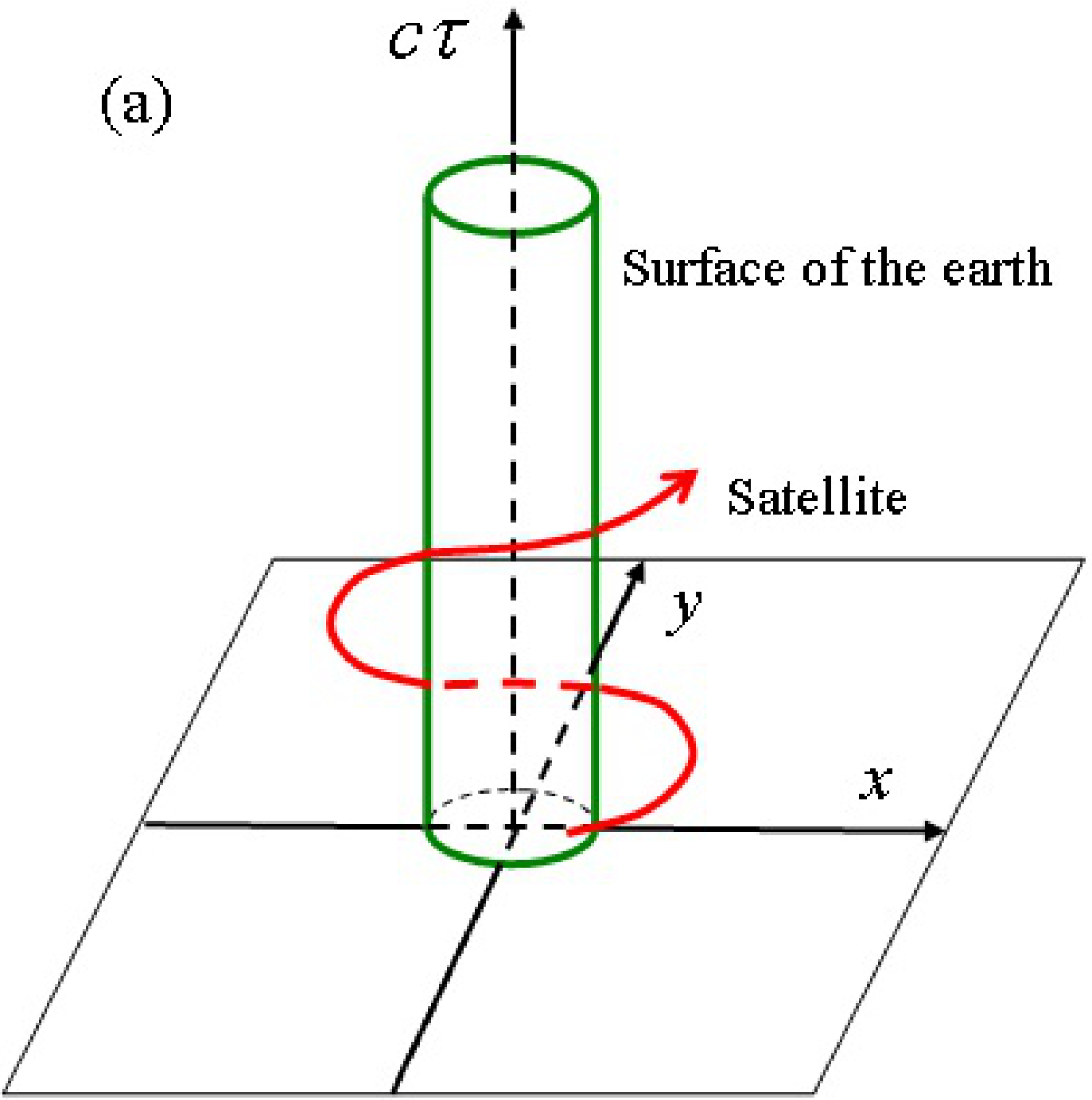}}
\subfigure{\includegraphics[width=4.5cm,height=5cm]{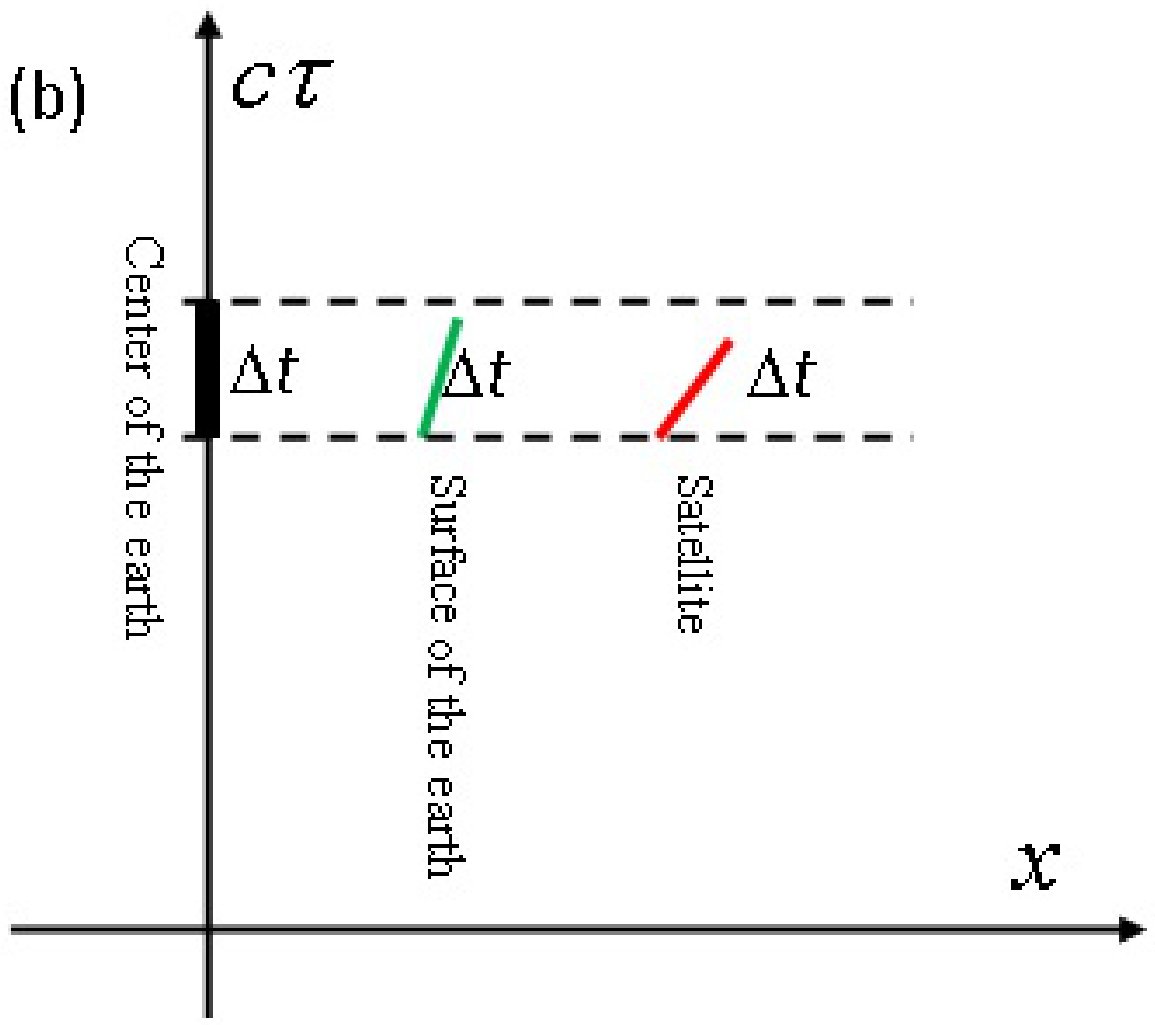}}
\caption{(a) World lines of a satellite (red curve) and  the earth surface (cylinder) at rest; (b) World lines taking account of the earth's spin.
}
\end{figure}
That is to say, the clock rest on a satellite goes slower by a factor of $\sqrt{1 - {GM\over rc^2}}$  than a clock rest on the earth.

Taking account of the spin of the earth, compare the proper time $\Delta \tau_1$ elapsed on the surface of the earth and $\Delta \tau_2$ on the satellite in the same time interval $\Delta t$ as shown in Fig.5(b). The geometry gives 
\begin{align}
\Delta\tau_i & =\Delta t \sqrt{1 - v_i^2/c^2}\label{tau1} 
\end{align}
where $v_i, i=1,2$ are velocities of the earth's surface and the satellite. The two proper times have a relative deviation ${(\Delta\tau_1 -\Delta\tau_2 )/\Delta\tau_1} $. 
This is the correction originating from the special theory of relativity. When $v_1$ is negligible to $v_2$ this result approaches to \eqref{sat}.

Now we consider the effect of gravity. A static and vacuum gravitational field outer of a mass $M$ is determined by the Schwarzschild metric\cite{Foster}
\begin{align}
c^2d\tau^2 = c^2 f(r)dt^2 - f(r)^{-1} dr^2 - r^2 d^2\Omega
\end{align}
where $f(r) = 1- {2GM\over c^2 r}$ and $d^2\Omega = d^2\theta + \sin^2\theta d^2\phi$. For a circular motion of a satellite $ dr =d\theta = 0, \theta = \pi/2$ and $d\phi = \omega dt$. Thus finally one has 
\begin{align} 
d\tau =\sqrt{f(r)- v^2/c^2} dt
\end{align}
This equation is similar to \eqref{tau1} but a new correction from the gravity enters into the function $f(r)$ in the above equation. This is the total relativistic effect of the satellite.   

This effect affects significantly the locating accuracy of a GPS system and has to be corrected on site to the clocks on satellites. 

\section{Summary}

A conventional space-time diagram is $r-ct$ one, which satisfies the Minkowski geometry. This geometry conflict the intuition from the Euclid geometry. In this work an Euclid space-time diagram is proposed to describe relativistic world lines with an exact Euclid geometry. The relativistic effects such as the dilation of moving clocks, the contraction of moving length, and the twin paradox can be geometrically expressed in the Euclid space-time diagram. It is applied to the case of a satellite clock to correct the gravitational effect. It is found that this Euclid space-time diagram is much more intuitive than the conventional space-time diagram.


\end{document}